\documentclass[prb,aps,twocolumn,superscriptaddress,floatfix,showpacs,longbibliography]{revtex4-2}
\usepackage{graphicx}
\usepackage{dcolumn}
\usepackage{bm}
\usepackage{upgreek}
\usepackage{natbib}
\usepackage[normalem]{ulem}
\usepackage{amsmath,amssymb}
\usepackage{dsfont}
\usepackage[english]{babel}
\usepackage{color}

\begin{document}

\title{Moir{\'e} exciton condensate: nonlinear Dirac point, broken-symmetry
Bloch waves and unusual optical selection rules}

\author{Sha Deng}
\affiliation{Guangdong Provincial Key Laboratory of Quantum Engineering and
Quantum Materials, School of Physics and Telecommunication Engineering, South
China Normal University, Guangzhou 510006, China}
\affiliation{Guangdong-Hong Kong Joint Laboratory of Quantum Matter, Frontier
Research Institute for Physics, South China Normal University, Guangzhou
510006, China}

\author{Yichen Chu}
\affiliation{Guangdong Provincial Key Laboratory of Quantum Engineering and
Quantum Materials, School of Physics and Telecommunication Engineering, South
China Normal University, Guangzhou 510006, China}
\affiliation{Guangdong-Hong Kong Joint Laboratory of Quantum Matter, Frontier
Research Institute for Physics, South China Normal University, Guangzhou
510006, China}

\author{Qizhong Zhu}
\email{qzzhu@m.scnu.edu.cn}
\affiliation{Guangdong Provincial Key Laboratory of Quantum Engineering and
Quantum Materials, School of Physics and Telecommunication Engineering, South
China Normal University, Guangzhou 510006, China}
\affiliation{Guangdong-Hong Kong Joint Laboratory of Quantum Matter, Frontier
Research Institute for Physics, South China Normal University, Guangzhou
510006, China}

\date{\today}

\begin{abstract}
  Moir{\'e} exciton features tunable Dirac dispersion and spatially
  dependent optical selection rules. With the long lifetime due to its interlayer nature, it
  is promising to realize a Bose-Einstein condensation of moir\'e excitons. Here we study the properties of moir{\'e} exciton condensate within the mean-field theory, with special focus on exciton-exciton interaction effect on the nonlinear Bloch band and Bloch waves of 
  exciton condensate in moir\'e potential. We find the nonlinear dispersion of the moir{\'e} exciton condensate exhibits complex loop structure induced by exciton-exciton interaction. A nonlinear Dirac cone emerges at $\boldsymbol{\Gamma}$ point of the Bloch band, with a three-fold
  degenerate nonlinear Dirac point.
  Each degenerate Bloch wave at nonlinear Dirac point spontaneously breaks $C3$ rotational symmetry of the moir\'e potential, and themselves differ by $C3$ rotations.
   Since they reside in the light cone, these nontrivial Bloch band structure and broken-symmetry Bloch waves can be experimentally detected by examining light emission from those
   states. Symmetry breaking of Bloch states implies
  unusual optical selection rules compared with single exciton case: moir\'e exciton condensate at $\boldsymbol{\Gamma}$ point emits light with all three components of polarization instead of only left or right circular polarization. 
  We further propose that, by applying in-plane electric field on one layer
  to drive an initially optically dark exciton condensate towards light cone, the light
  polarization of final states as well as their dependence on field direction serves as the smoking gun for experimental observation.
\end{abstract}

{\maketitle}

\section{introduction}

By stacking two identical or similar two-dimensional materials with a small twist angle, one can realize various forms of moir\'e pattern. As a prototype of twisted bilayer materials,
twisted bilayer graphene has become an intriguing platform for exploring a variety
of interesting correlated and topological phases \cite{bistritzer_Moire_2011,cao_Correlated_2018,cao_Unconventional_2018,lu_Superconductors_2019,
kerelsky_Maximized_2019,jiang_Charge_2019,xie_Spectroscopic_2019,song_All_2019}.
Subsequently, other bilayer systems consisting of various
two-dimensional semiconductors have been proposed to host interesting physics, with
twisted bilayer transition metal dichalcogenides (TMDs) as an especially attractive candidate. 
On the one hand, similar topological and correlated phases
are also present in twisted bilayer TMDs \cite{wu_Hubbard_2018,wu_Topological_2019,kennes_Moire_2021,tang_Simulation_2020,xu_Correlated_2020,shimazaki_Strongly_2020,regan_Mott_2020,zhou_Bilayer_2021,smolenski_Signatures_2021}, which is an interesting subject to study. On the
other hand, twisted bilayer TMDs also presents an unique platform for exploring the interplay between
excitonic physics and moir\'e pattern (see, e.g., the review papers \cite{wilson_Excitons_2021,huang_Excitons_2022}). In particular, the interesting concept of moir\'e exciton 
is proposed \cite{yu_Moire_2017} and experimentally confirmed afterwards \cite{seyler_Signatures_2019,tran_Evidence_2019,jin_Observation_2019,
alexeev_Resonantly_2019,jin_Identification_2019,wang_Diffusivity_2021,forg_Moire_2021}.
 As a type of interlayer exciton, moir{\'e} exciton is trapped by periodic potential
 generated by moir\'e superlattice, whose optical selection rule
 is spatially modulated \cite{yu_Moire_2017,wu_Theory_2018}. 
Interestingly, the dispersion of moir\'e exciton resembles that of
 massive Dirac fermion, which can be tuned by interlayer
electric field \cite{yu_Moire_2017,yu_Electrically_2020}. 

With the advantage of interlayer exciton, it is promising that Bose-Einstein condensation of moir\'e exciton may occur at
low enough temperature and high enough density \cite{lagoin_Key_2021}. 
Actually, much progress has been made towards the realization of a Bose-Einstein condensate of
interlayer excitons \cite{wang_Evidence_2019,sigl_Signatures_2020}.
The lifetime of moir\'e exciton can reach the order of 100 ns at certain twist angles \cite{choi_Twist_2021}, which is favorable for the realization of a moir\'e exciton condensate. Although at ground state the moir\'e exciton
condensate is optically dark \cite{lagoin_Key_2021,remez_Leaky_2022}, it can emit light with the help of exciton-exciton interaction \cite{remez_Leaky_2022}
or phonon scattering \cite{forg_Moire_2021}. In fact, there are other ways to turn the condensate
bright, e.g., by driving the exciton toward the optically bright $\boldsymbol{\Gamma}$ point,
which can be realized by applying an in-plane electric field on only one layer.
Similar methods have been routinely used in the community of cold atomic gases to prepare a
condensate with nonzero quasimomentum \cite{bendahan_Bloch_1996}. 
To have a clear understanding of this process, one needs the
knowledge of nonlinear Bloch band and Bloch wave of moir\'e exciton condensate. 
Similar issue has been extensively studied
in cold atoms, and intriguing phenomenon such as
swallowtail or loop structure of nonlinear Bloch dispersions has been discovered \cite{wu_Bloch_2002,machholm_Band_2003}.
In particular, the nonlinear Bloch band of an atomic condensate in a honeycomb
lattice potential was studied by Chen and Wu \cite{chen_BoseEinstein_2011}. There is
also report on the  
nonlinear Bloch band of exciton-polariton condensates \cite{chestnov_Nonlinear_2016}. However, what really makes moir\'e exciton condensate special is that, the interesting features related with loop structure around $\boldsymbol{\Gamma}$ point lies
within light cone, and thus can be probed directly in experiments from light emission. So this system provides an ideal platform for investigating the interplay
between nonlinear Bloch states and light emission of exciton.

In this paper, we aim to explore the properties of nonlinear
Bloch band and Bloch states of moir{\'e} exciton condensate, taking into account the effect of exciton-exciton interaction in the framework of mean-field theory. We find that
exciton-exciton interaction has a dramatic effect on the nonlinear dispersion of exciton condensate, by
introducing additional loop structure and causing symmetry breaking of Bloch waves. Those loop structure of Bloch band renders the formation of a nonlinear Dirac cone \cite{bomantara_Nonlinear_2017}. In addition, the nonlinear Dirac point at $\boldsymbol{\Gamma}$ point has
three-fold degeneracy, with each degenerate state breaking $C3$ rotational symmetry.
Associated with symmetry
change of Bloch waves, exciton-exciton interaction also induces a significant reduction of
Berry curvature of Bloch band compared with non-interacting case.
Our findings can be experimentally demonstrated by a remarkable
 phenomenon: starting from a dark exciton condensate at ground
state, e.g., at $-\boldsymbol{K}_m$ point of moir\'e mini-Brillouin zone (mini-BZ), by applying in-plane electric field to only one layer along
three different directions from $-\boldsymbol{K}_m$ to $\boldsymbol{\Gamma}$, 
the moir\'e exciton condensate will be driven to the three different degenerate
Bloch states, respectively. The light polarization at final state depends on the direction of the applied electric field.
Besides, Bloch waves at $\boldsymbol{\Gamma}$ point no longer emit left- or right-handed circularly
polarized light, but instead, light with $\sigma_+$, $\sigma_-$ and $z$ components.
Additionally, the reduction of Berry curvature of nonlinear Bloch band will be
manifested by a much weaker Hall effect compared
with single exciton case.

The rest of this paper is organized as follows. In Sec. \ref{sec2}, we present the theoretical framework
for studying the properties of moir\'e exciton condensate, and show the features of nonlinear Bloch band and symmetry of nonlinear Bloch waves. The optical
selection rules of those nonlinear Bloch waves is explored in Sec. \ref{sec3}. In Sec. \ref{sec4} we calculate the Berry curvature of nonlinear Bloch band with the presence of  exciton-exciton interaction. With these theoretical results, we propose an experimental scheme to test our
predictions, discuss the effect of exciton condensate in another spin species and make a summary in Sec. \ref{sec5}.

\section{nonlinear Bloch band and Bloch wave}
\label{sec2}

\begin{figure}
\centering
\includegraphics[width=0.95\linewidth]{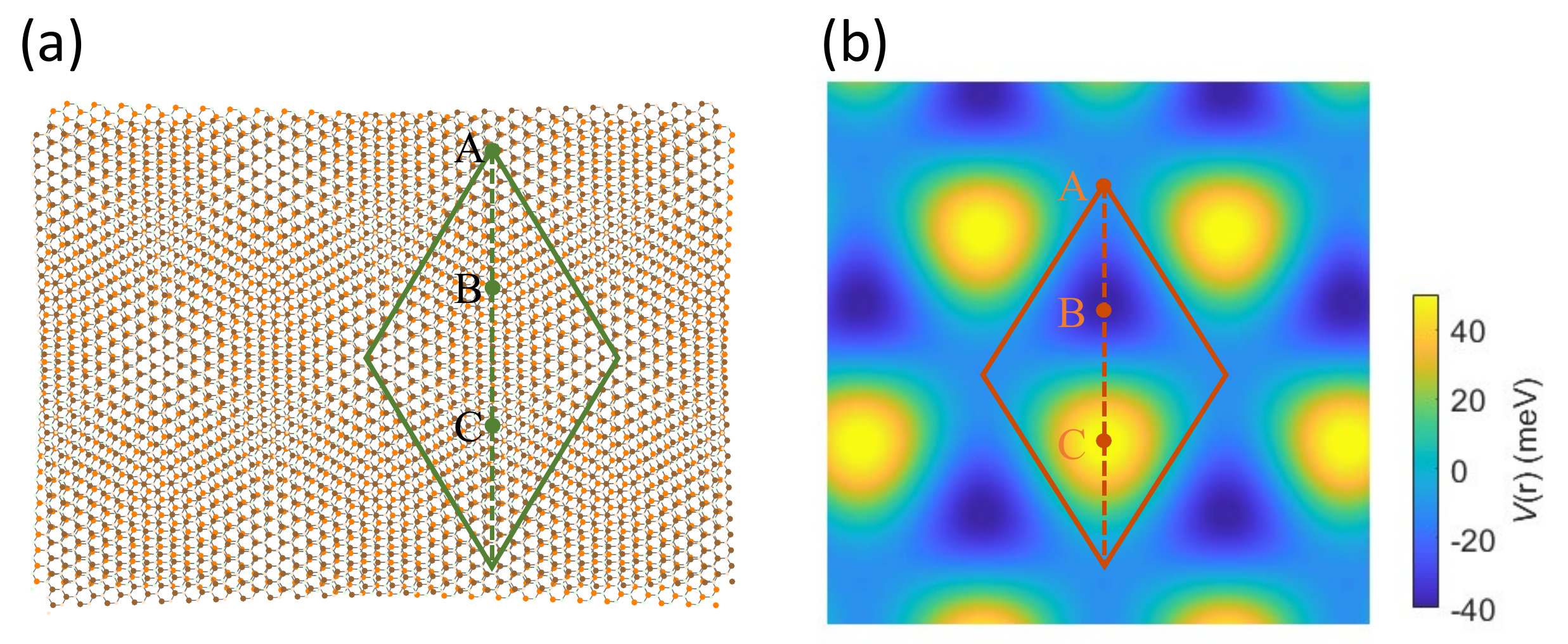}
\caption{(a) Stacking registry of moir\'e superlattice formed by twisted bilayer TMDs. The unit cell is indicated by the diamond shape, along with three high symmetry points A, B and C, where $C3$ rotational symmetry is respected. (b) A typical potential landscape that moir\'e exciton feels in the moir\'e superlattice. A, B and C points have 
the same meaning with (a). $V_0=-2.0-8.9i$ meV. \label{fig1}}
\end{figure}

A moir\'e superlattice can be formed by stacking bilayer TMDs with a small twist angle.
 Figure \ref{fig1}(a) shows
a typical moir\'e superlattice and its unit cell, where A, B and C denote
high symmetry locations that respect $C3$ rotational symmetry. Moir\'e exciton, with
electron and hole residing in different layers, feels a spatially varying potential
with the same periodicity, according to first-principles calculations \cite{guo_Shedding_2020}. The high symmetry
points in stacking registry and potential landscape
have a one-to-one correspondence, as indicated by same labels in fig. \ref{fig1}(a) and (b).
The moir\'e potential has a hexagonal lattice structure, with two local potential minima A and B. The energy difference between A and B can
be experimentally tuned by applying an out-of-plane electric field.

\begin{figure}
\centering
\includegraphics[width=0.99\linewidth]{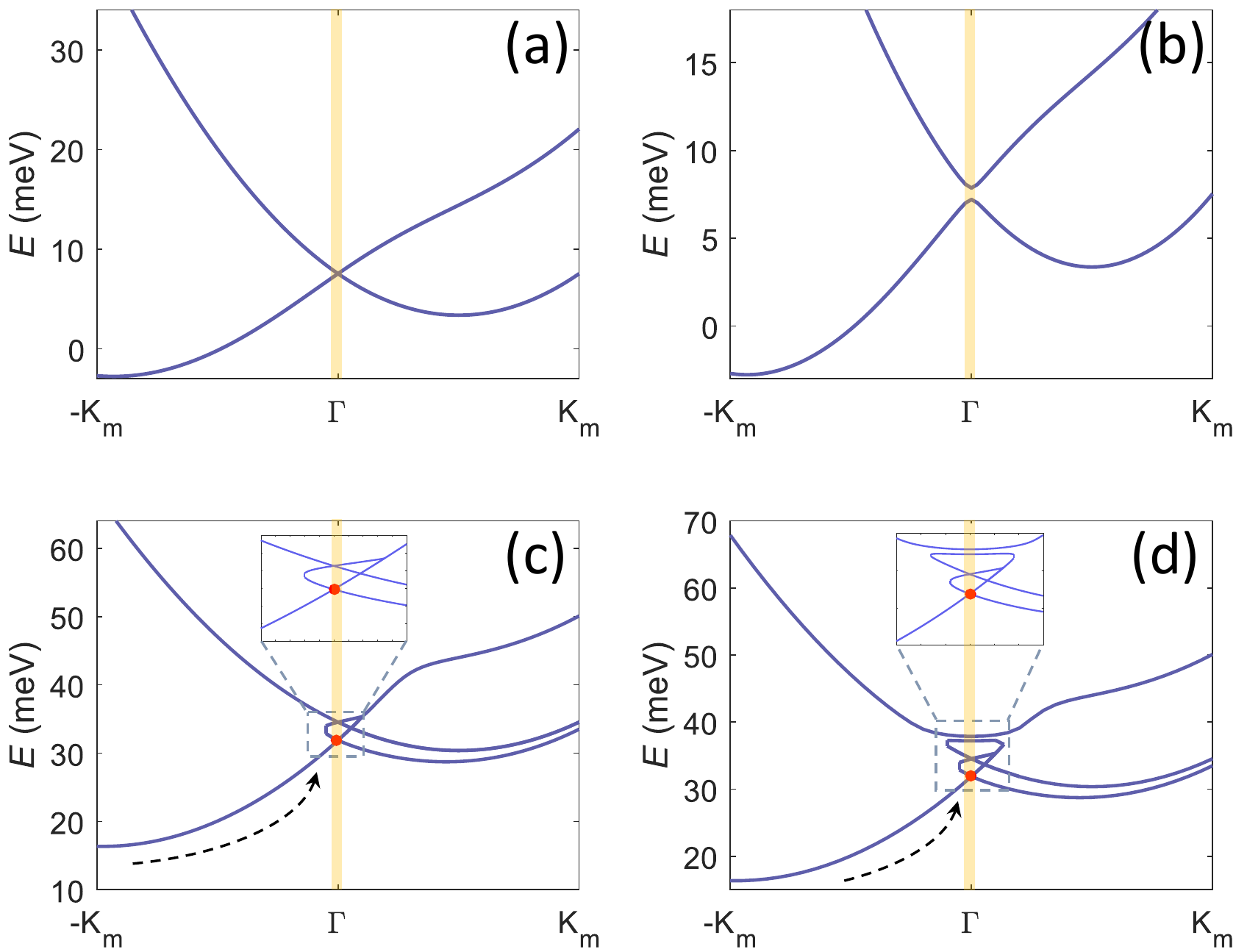}
\caption{Non-interacting/single moir\'e exciton dispersion for A-B inversion symmetric (a) and asymmetric (b) potentials. $V_0=9.1$ meV (a) and $V_0=-4.4-8i$ meV (b). (c) and (d) are nonlinear exciton dispersions $\mu(\boldsymbol{k})$ with interaction
corresponding to (a) and (b), respectively.  Red dots denote the nonlinear Dirac points relevant to the optical selection rules studied in the following. The orange shaded area indicates the location of light cone and the dashed line with arrow indicates the evolution of state from $-\boldsymbol{K}_m$ to $\boldsymbol{\Gamma}$. $g=18$ meV$\cdot a^2$ in (c) and (d). $a=3.6$ nm, corresponding to a twist angle of $\sim 5^{\circ}$.  \label{fig2}}
\end{figure}

The minimal potential model with the required
symmetry can be expanded in terms of three Fourier components, $V
(\boldsymbol{r}) = V_0 (e^{i\boldsymbol{b}_1 \cdot \boldsymbol{r}} +
e^{i\boldsymbol{b}_2 \cdot \boldsymbol{r}} + e^{i\boldsymbol{b}_3 \cdot
\boldsymbol{r}}) + V_0^* (e^{- i\boldsymbol{b}_1 \cdot
\boldsymbol{r}} + e^{- i\boldsymbol{b}_2 \cdot \boldsymbol{r}} + e^{-
i\boldsymbol{b}_3 \cdot \boldsymbol{r}})$, where $\boldsymbol{b}_1=\frac{2\pi}{3a}(0,-2)$ and
$\boldsymbol{b}_2=\frac{2\pi}{3a}(\sqrt{3},1)$ are two reciprocal lattice vectors, $\boldsymbol{b}_3=-\boldsymbol{b}_1
-\boldsymbol{b}_2$, with $a$ being the lattice constant of moir\'e potential.
We first only consider moir\'e excitons formed by electron and hole both in $\boldsymbol{K}$ valley of monolayer TMD, hereafter referred to as spin-up exciton. The result for $-\boldsymbol{K}$ valley exciton or spin-down exciton can be obtained by taking time reversal operation.
The momentum of $\boldsymbol{K}_m$ valley in moir\'e mini-BZ is related to the momentum difference between electron $\boldsymbol{K}_e$ and hole $\boldsymbol{K}_h$ valley differing by a twist, namely, $\boldsymbol{K}_m=\boldsymbol{K}_e-\boldsymbol{K}_h$.
Bloch bands of spin-up moir{\'e} excitons in this potential 
are shown in
figs. \ref{fig2}(a) and (b). The dispersions resemble that of a Dirac fermion,
 whose mass is
determined by the potential energy difference between A and B points. The dispersion of
spin-down moir\'e exciton is its time reversal partner.

Given the dispersion of a single moir{\'e} exciton, it is evident that 
the ground state of many excitons
is a moir{\'e} exciton condensate at $-\boldsymbol{K}_m$ point, which is optically dark as only those
excitons with momentum in the light cone (indicated by orange shaded areas in fig. \ref{fig2}) around $\boldsymbol{\Gamma}$ point can recombine.
However, the condensate can emit light with the help of
exciton-exciton scattering \cite{remez_Leaky_2022} or by absorption of phonons \cite{forg_Moire_2021}. Here we point out another mechanism for light emission, namely, 
by driving the exciton condensate
towards optically bright $\boldsymbol{\Gamma}$ point, with an electric field applied only in one layer.
Assuming the electric field is weak and the above process is adiabatic, the moir{\'e} exciton
condensate will follow the
nonlinear dispersion from $-\boldsymbol{K}_m$ to $\boldsymbol{\Gamma}$ (shown by dashed line with arrow in figs. \ref{fig2}(c) and (d)), which is the central issue we will address in the following. 

For simplicity, we assume a contact interaction between moir\'e excitons, and treat
the interaction within the framework of mean-field theory, as commonly done in the study of atomic condensate. For spin-up moir\'e exciton, a condensate at ground state with momentum $-\boldsymbol{K}_m$ can be experimentally prepared by first exciting intralayer excitons with circularly polarized light, followed by fast interlayer charge transfer.
 The stationary Gross-Pitaevskii
equation (GPE) that takes into account the effect of both moir\'e potential and
exciton-exciton interaction reads \cite{pethick_Bose_2008}
\begin{equation}
\mu \psi = \left[ \frac{(\hat{P}  - \boldsymbol{A})^2}{2 m} + V (\boldsymbol{r}) + g | \psi |^2
   \right] \psi,\label{eq1}
\end{equation} 
where $\psi$ is the condensate wave function, $V (\boldsymbol{r})$ is the moir\'e periodic potential, $g$ is the strength of contact interaction between excitons,
 and $\mu$ is the chemical potential. 
 The exciton-exciton interaction energy in mean-field theory
plays the role of an
effective potential, self-consistently determined by the nonlinear Bloch wave
function.
 We here introduce a constant vector potential $\boldsymbol{A}=-\boldsymbol{K}_m$ to account for the
 nonzero center-of-mass momentum of excitons at dispersion minimum \cite{yu_Moire_2017}. The non-interacting dispersion resembles that of graphene, except with Dirac cone shifted
 to $\boldsymbol{\Gamma}$ point.
 The Hamiltonian of spin-down moir\'e exciton is time reversal of Eq. \ref{eq1}. 
 The above GPE can be solved
by expanding the nonlinear Bloch waves in terms of plane waves, $\psi_{\boldsymbol{k}}(\boldsymbol{r}) = \sum_j
C_j (\boldsymbol{k}+\boldsymbol{G}_j) e^{i (\boldsymbol{k}+\boldsymbol{G}_j) \cdot \boldsymbol{r}}$,
with reciprocal lattice vectors $\boldsymbol{G}_j=m\boldsymbol{b}_1
+n\boldsymbol{b}_2$.
One then solves a set of nonlinear equations in terms of these expansion coefficients
$C_j (\boldsymbol{k}+\boldsymbol{G}_j)$, with the constrain of wave function normalization. The nonlinear dispersion $\mu(\boldsymbol{k})$ as
well as wave function can be subsequently determined. 

For comparison, we first show the non-interacting dispersion of moir{\'e} exciton,
 i.e., $g = 0$, in figs. \ref{fig2}(a) and (b). Depending on whether the moir{\'e}
potential at A and B point has inversion symmetry or not, the exciton dispersion 
resembles that of a
massless or massive Dirac fermions, with gap $\delta$. The potential energy difference between
A and B can be tuned by an out-of-plane electric field. In the
asymmetric case, the non-degenerate Bloch wave at $\boldsymbol{\Gamma}$ point has $C3$ rotational symmetry, indicating that exciton at $\boldsymbol{\Gamma}$ point should couple to circularly polarized light.

Much more interestingly, with the presence of
exciton-exciton interaction, i.e., $g\neq 0$, the
exciton dispersion and Bloch wave function can be strongly modified. The nonlinear dispersion $\mu (\boldsymbol{k})$ is plotted in figs. \ref{fig2}(c) and
(d) along the high symmetry path from $-\boldsymbol{K}_m$ to $\boldsymbol{K}_m$. The overall nonlinear
dispersion in two-dimensional moir\'e mini-BZ
 has $C3$ rotational symmetry. It is evident that the nonlinear
dispersion exhibits new structures absent in the non-interacting case. There are
extra energy levels, since nonlinear
equations can have more solutions that their linear counterpart. Similar loop or swallowtail structure has been reported in
nonlinear Bloch band of atomic condensate in optical lattices \cite{chen_BoseEinstein_2011}. 
Besides, a nonlinear Dirac cone \cite{bomantara_Nonlinear_2017} emerges
due to exciton-exciton interaction, with Dirac point indicated by red dots in figs. \ref{fig2}(c) and (d).
 In particular, when the
single particle gap vanishes ($\delta=0$), arbitrarily weak exciton-exciton interaction will
cause the appearance of loop structure. On the other hand, if $\delta$ is
finite, this occurs only when the strength of the exciton-exciton
interaction $g$ exceeds a threshold that depends on $\delta$.

\begin{figure}
\centering
\includegraphics[width=0.99\linewidth]{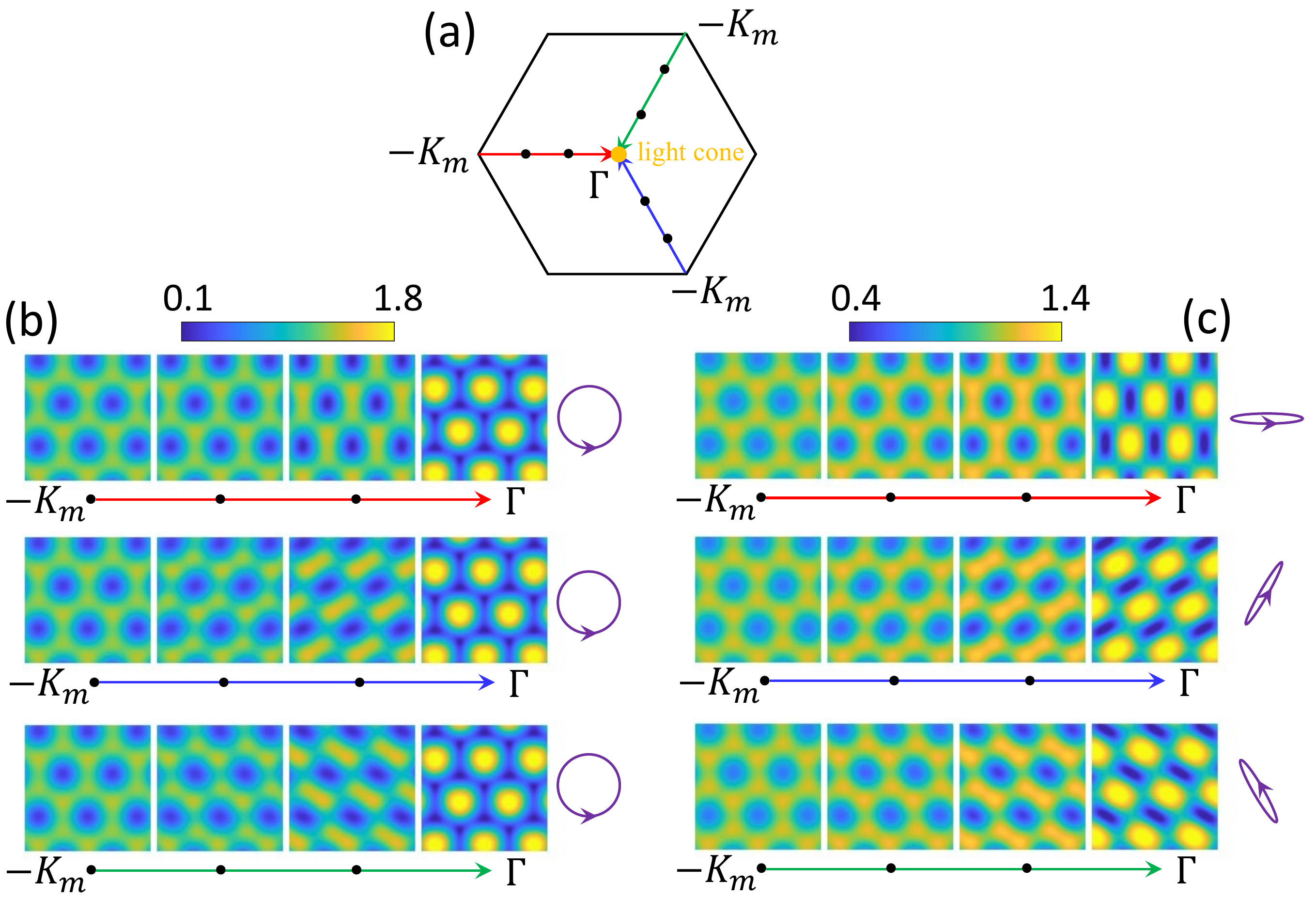}
\caption{(a) Three different paths from $-\boldsymbol{K}_m$ to $\boldsymbol{\Gamma}$ in moir\'e mini-BZ, indicated by red, blue and green arrows, respectively. The central orange disk shows the light cone.
(b) Amplitude of Bloch wave function $|\psi_{\boldsymbol{k}}(\boldsymbol{r})|$ at four points along those three paths in the non-interacting (b) and interacting case (c). The light polarization emitted at $\boldsymbol{\Gamma}$ is indicated by polarization ellipse ($z$ component of polarization is omitted for simplicity). Parameters in (b) and
(c) are the same as figs. \ref{fig2}(b) and (d), respectively. \label{fig3}}
\end{figure}

Besides the peculiar structure of nonlinear Bloch band,
additional degeneracy is
also brought by interaction. After inspecting the Bloch wave functions, we find that in contrast to the 
non-interacting case where Bloch wave at $\boldsymbol{\Gamma}$ point is unique for nonzero single particle gap,
 there is
three-fold degeneracy at the newly emergent nonlinear Dirac point (red dots in figs. \ref{fig2}(c) and (d)). 
Of particular interest is the symmetry of degenerate states at nonlinear Dirac point,
 because these states can be directly coupled with light.
We find that
 at $\boldsymbol{\Gamma}$ point, the $C3$ rotational symmetry
of Bloch state is spontaneously broken, and the three
degenerate Bloch waves at the nonlinear Dirac point differ by $C3$ rotation with respect to each other. 
This is allowed since these three states contribute different forms of effective
potential, i.e., $g|\psi(\boldsymbol{r})|^2$ and correspond to
three different effective Hamiltonians.
To illustrate how these three degenerate states are
connected with the unique state at $-\boldsymbol{K}_m$ point, we show the wave function along three different paths from $-
\boldsymbol{K}_m$ to $\boldsymbol{\Gamma}$, in fig. \ref{fig3}(c). In contrast to the non-interacting case,
where the initial and final states are both unique (fig. \ref{fig3}(b)), in the interacting
case, the initial states are same, but final states are three different
Bloch states. The symmetry breaking of degenerate states at
$\boldsymbol{\Gamma}$ point brings dramatic change to the
light polarization of those states, as shown by polarization ellipses in
fig. \ref{fig3}(c). Their optical selection rules will be
discussed in Sec. \ref{sec3} and can be
experimentally tested using the scheme proposed in Sec. \ref{sec5}.
Note that $C3$ rotational symmetry of exciton wave function can also be broken by introducing strain \cite{zheng_Twist_2021}, which however is not spontaneous as the symmetry of Hamiltonian is changed. 

\section{optical selection rules at nonlinear Dirac point}
\label{sec3}

The symmetry of degenerate Bloch states at $\boldsymbol{\Gamma}$ point is directly reflected in
their optical selection rules. In non-interacting case, for finite single particle gap, Bloch state at $\boldsymbol{\Gamma}$ point has $C3$ rotational symmetry, and can only
couple to photon with circular polarization.

In the interacting case, the nonlinear Dirac point has three-fold degeneracy. 
Each degenerate Bloch wave spontaneously breaks the $C3$
rotational symmetry, accompanied by the
change of optical selection rules. We calculate the optical dipole of those
states by considering spin-up exciton at the three main light cones \cite{yu_Anomalous_2015,yu_Electrically_2020}, $\alpha_{\boldsymbol{k}\uparrow}$,
$\beta_{\boldsymbol{k}\uparrow}$ and $\gamma_{\boldsymbol{k}\uparrow}$, corresponding to
 $\boldsymbol{K}_m$, $C_3\boldsymbol{K}_m$ and $C_3^2\boldsymbol{K}_m$ momentum components in the expansion of nonlinear Bloch waves, respectively. Spin-up excitons at three main light
cones have the following optical dipoles respectively \cite{yu_Electrically_2020},
\begin{align}
  D_{\alpha \uparrow} & =  D_+ \boldsymbol{e}_+^{\ast} + D_-
  \boldsymbol{e}_-^{\ast} + D_z \boldsymbol{e}_z,\nonumber\\
  D_{\beta \uparrow} & =  D_+ \boldsymbol{e}_+^{\ast} + e^{- i 2 \pi / 3} D_-
  \boldsymbol{e}_-^{\ast} + e^{i 2 \pi / 3} D_z \boldsymbol{e}_z,\nonumber\\
  D_{\gamma \uparrow} & =  D_+ \boldsymbol{e}_+^{\ast} + e^{i 2 \pi / 3} D_-
  \boldsymbol{e}_-^{\ast} + e^{- i 2 \pi / 3} D_z \boldsymbol{e}_z ,
\end{align}
where $\boldsymbol{e}_\pm=(\boldsymbol{e}_x\pm i\boldsymbol{e}_y)/\sqrt{2}$, 
$D_+$, $D_-$ and $D_z$ can be obtained from first-principles calculations.
For a general state expressed as superposition of three main light cones, $|
A_{\boldsymbol{k}\uparrow} \rangle = c_1 | \alpha_{\boldsymbol{k}\uparrow} \rangle
+ c_2 | \beta_{\boldsymbol{k}\uparrow} \rangle + c_3 |
\gamma_{\boldsymbol{k}\uparrow} \rangle$, its optical
dipole can be calculated using the formula $D_{A\uparrow}=c_1 D_{\alpha\uparrow}+c_2 D_{\beta\uparrow}
+c_3 D_{\gamma\uparrow}$, with $c_1$, $c_2$ and $c_3$ being the expansion coefficients. For the
three degenerate states at $\boldsymbol{\Gamma}$ point in fig. \ref{fig3}(c),
 they have optical dipoles as follows respectively,
 \begin{align}
 D_1\approx & 1.16D_+\boldsymbol{e}_+^{\ast}+1.24D_-\boldsymbol{e}_-^{\ast}
 +0.32D_z\boldsymbol{e}_z,\nonumber\\ 
 D_2\approx & 1.16D_+\boldsymbol{e}_+^{\ast}+1.24e^{- i 2\pi/3}D_-\boldsymbol{e}_-^{\ast}+0.32e^{i 2\pi/3}D_z\boldsymbol{e}_z,\nonumber\\
 D_3\approx & 1.16D_+\boldsymbol{e}_+^{\ast}+1.24e^{i 2\pi/3}D_-\boldsymbol{e}_-^{\ast}
 +0.32e^{-i 2\pi/3}D_z\boldsymbol{e}_z.
\end{align} 
Note that their optical dipoles have all three components of polarization, namely,
left-/right-handed circular polarization as well as polarization in $z$ direction. Light polarization of the states at $\boldsymbol{\Gamma}$ point serves as the smoking gun for the experimental detection of their symmetry.

\begin{figure}
\centering
\includegraphics[width=0.9\linewidth]{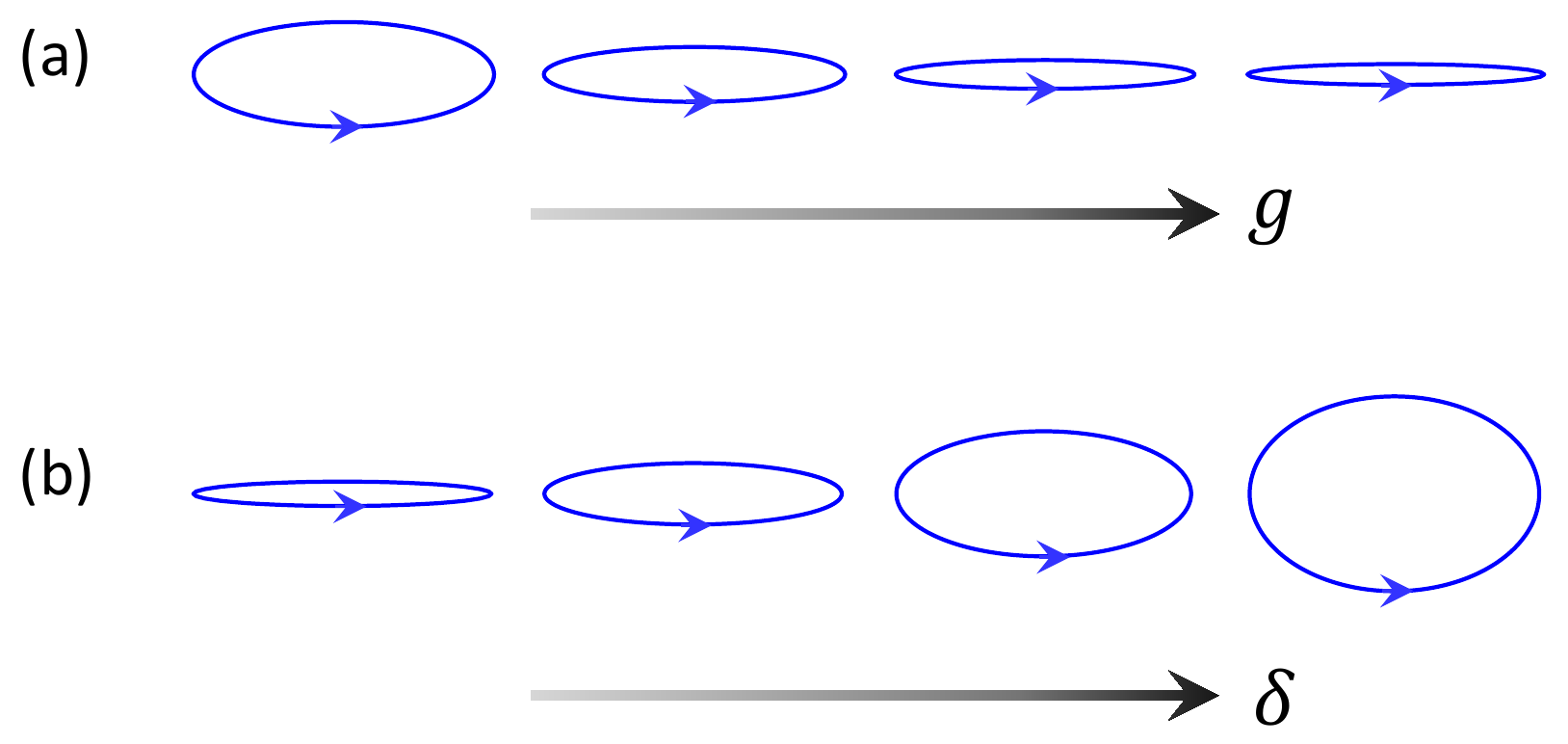}
\caption{(a) Evolution of polarization ellipse for one Bloch wave at nonlinear Dirac point with the increase of interaction $g$ at fixed gap $\delta$. $\delta=1.9$ meV and $g=6, 10, 16, 20$ meV$\cdot a^2$ from left to right.
(b) Evolution of polarization ellipse with the increase of $\delta$ at fixed $g$. $g=18$ meV$\cdot a^2$ and
$\delta=0.01, 5.1, 9.9, 15.3$ meV from left to right. In the calculations, a reasonable assumption $D_+\approx D_-$ is made. \label{fig4}}
\end{figure}

For the purpose of
easy illustration, we now only consider the in-plane components of optical dipole, and show the polarization ellipses of those degenerate states in fig. \ref{fig3}(c).
Their optical dipoles and associated polarization ellipses change with the single particle gap $\delta$ and exciton-exciton interaction $g$. 
With fixed $\delta$,
the ellipticity increases with $g$. On the other hand, with fixed $g$, the ellipticity decreases with $\delta$. This trend is shown in fig. \ref{fig4}. In other words, the more the interaction $g$ exceeds the gap-dependent threshold $g_c(\delta)$, the larger the ellipticity is.

\section{Berry curvature of nonlinear Bloch band}
\label{sec4}

\begin{figure}
\centering
\includegraphics[width=0.85\linewidth]{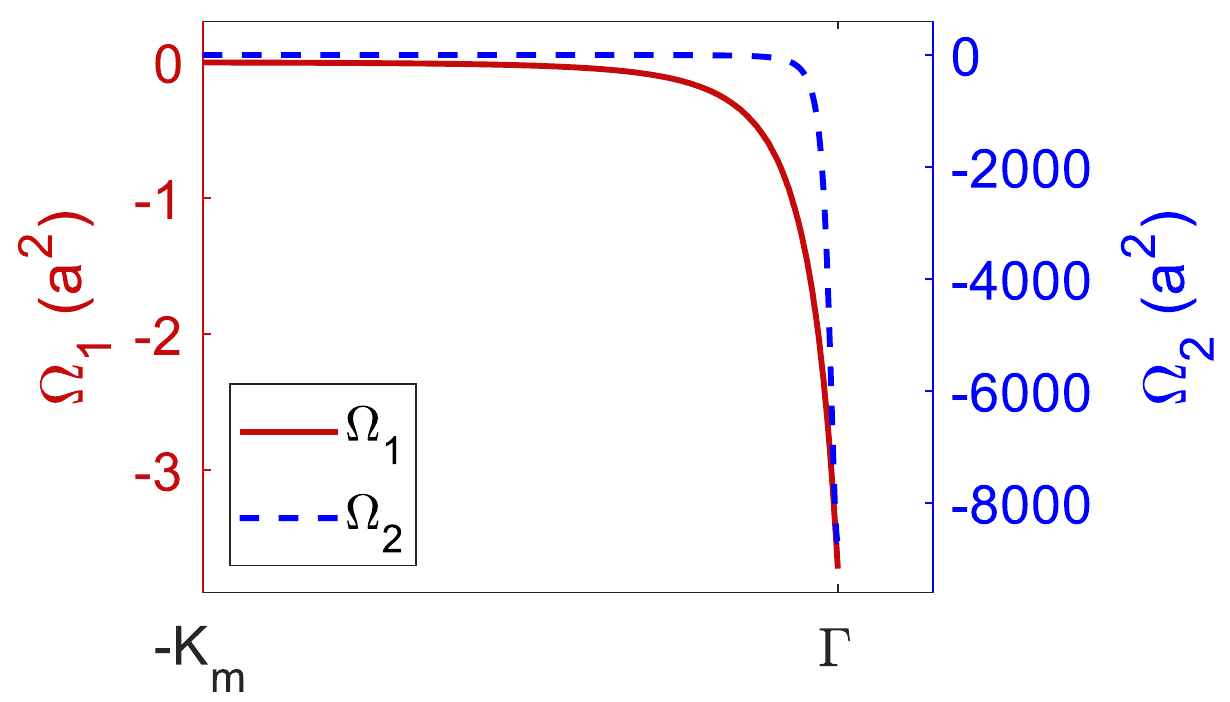}
\caption{Berry curvature of the lowest Bloch band along the path $-\boldsymbol{K}_m$ to $\boldsymbol{\Gamma}$
for the interacting (left vertical axis, $\Omega_1$) and non-interacting case (right vertical axis, $\Omega_2$). The parameters are the same as figs. \ref{fig2}(d) and (b), respectively. \label{fig5}}
\end{figure}

Besides the symmetry change of wave function, interaction also has dramatic
effect on Berry curvature. We calculate the Berry curvature of the lowest nonlinear Bloch band \cite{xiao_Berry_2010},
\begin{equation}
\Omega_n(\boldsymbol{k})=\left\langle\frac{\partial u_{n\boldsymbol{k}}}{\partial\boldsymbol{k}}\bigg|
\times\bigg|\frac{\partial u_{n\boldsymbol{k}}}{\partial\boldsymbol{k}}\right\rangle,
\end{equation}
where $u_{n\boldsymbol{k}}$ is the periodic part of nonlinear Bloch wave function at band $n$ and
momentum $\boldsymbol{k}$.
The nonlinear Berry curvature is numerically evaluated using the method in Ref. \cite{fukui_Chern_2005} and the result is shown 
in fig. \ref{fig5}. Here we are only interested in the Berry curvature along the
path from $-\boldsymbol{K}_m$ to $\boldsymbol{\Gamma}$, as it affects the transverse motion of moir\'e exciton condensate when dragged from $-\boldsymbol{K}_m$ to $\boldsymbol{\Gamma}$.
One can see that the
magnitude of Berry curvature in the interacting case is orders of
magnitude smaller than its non-interacting counterpart. This is reasonable since in the
interacting case the wave function changes more smoothly along the path $-\boldsymbol{K}_m$ to
$\boldsymbol{\Gamma}$, as shown in fig. \ref{fig3}(c). The dramatic reduction of Berry curvature
can be tested by measuring the transverse drift of moir{\'e} exciton condensate
when dragged by an in-plane electric field. The moir\'e exciton condensate will exhibit
a much weaker Hall effect compared with single exciton case. Note that,
in the calculation of Berry curvature of a nonlinear Bloch band, although there is band crossing at nonlinear Dirac point,
the degenerate states actually live in different Hilbert spaces which all have gapped spectrum, and thus we can still define an abelian Berry curvature as in the non-degenerate case.

\section{discussion and conclusion}
\label{sec5}

Based on our theoretical results, we predict
a prominent experimental signatures of moir\'e exciton condensate: non-circular polarization of light emission at $\boldsymbol{\Gamma}$ with dependence on
field direction. In experiments, intralayer exciton can be first
excited by circularly polarized light, followed by fast interlayer charge transfer. When (quasi-)equilibrium is reached with low temperature and high density, a moir\'e exciton condensate at $-\boldsymbol{K}_m$ point
is prepared.
By applying an electric field on one layer,
 the exciton condensate will be dragged in momentum space
from $-\boldsymbol{K}_m$ to $\boldsymbol{\Gamma}$ point along chosen path adiabatically if
the electric field is weak \cite{bendahan_Bloch_1996}. When the exciton condensate arrives at $\boldsymbol{\Gamma}$
point, it will produce photoluminescence with all three components of polarization, in stark contrast with the circular polarization in single exciton case. In addition, by applying electric field along three
different paths from $-\boldsymbol{K}_m$ to $\boldsymbol{\Gamma}$ (see fig. \ref{fig3}(c)), the light polarization at $\boldsymbol{\Gamma}$ is also different.
During the evolution process, one can also measure the transverse drift of exciton condensate \cite{rivera_Valleypolarized_2016}, which will be
much smaller than its single exciton counterpart.

\begin{figure}
\centering
\includegraphics[width=0.98\linewidth]{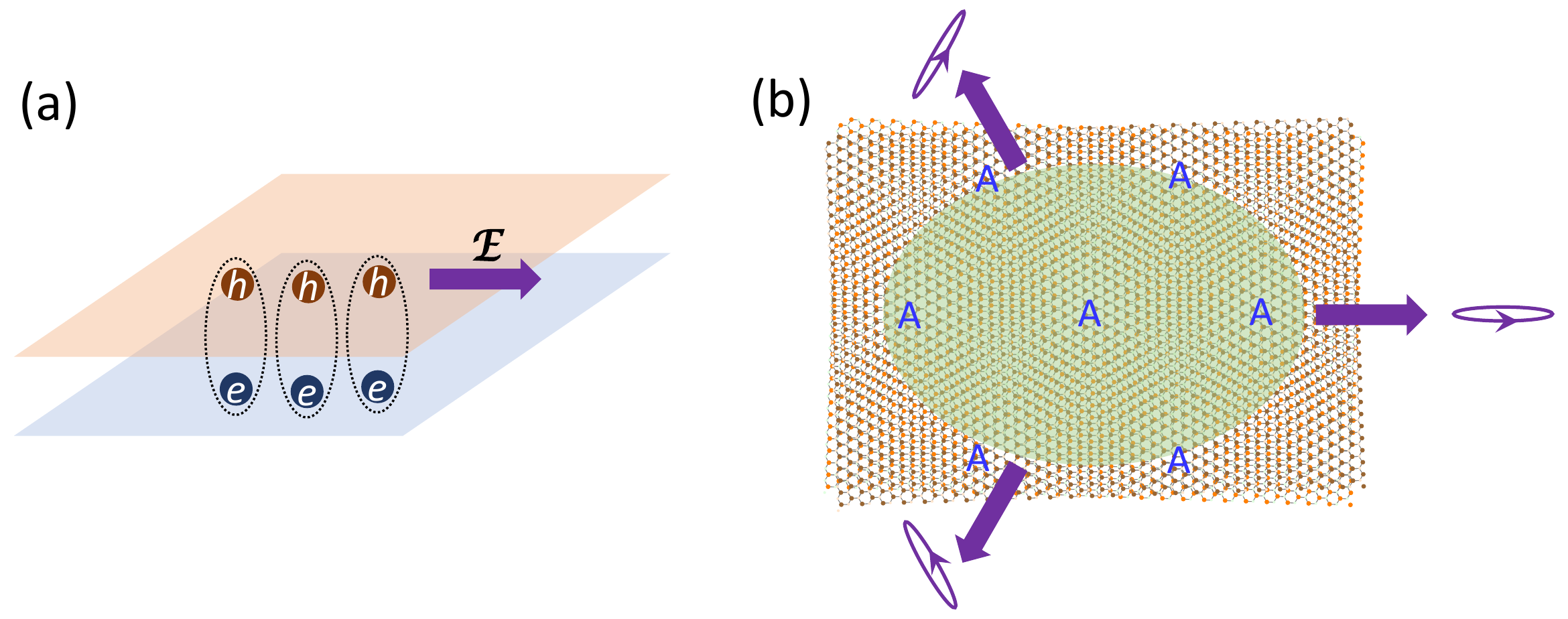}
\caption{(a) Moir\'e exciton condensate driven by in-plane electric field applied only
on one layer. (b) Top view of experimental sample with three purple arrows
indicating the directions of applied electric fields, inducing evolution from $-\boldsymbol{K}_m$ to $\boldsymbol{\Gamma}$ along three different paths. The different in-plane light polarization of final states are illustrated by polarization ellipse.
 ``A'' denotes high symmetry points in moir\'e superlattice same as fig. \ref{fig1}(a). \label{fig6}}
\end{figure}

In realistic experiments, a hBN spacer may be intercalated between the TMD monolayers, which can significantly reduce the strength of moir\'e potential.
This is however favorable for the realization of moir\'e exciton superfluid \cite{lagoin_Key_2021}. We have numerically reduced the strength of moir\'e potential by one order of magnitude, and find no qualitative change to the conclusion reached above.

We also examined the case where moir\'e exciton condensates with two spin species coexist.
The condensate mixture with two spin species can be described by a spinor GPE,
with the inter-species exciton-exciton interaction weaker than intra-species one \cite{yu_Moire_2017}. We numerically find that the nonlinear dispersion is further
splitted, and the detailed structure is much more complicated. 
Nevertheless, we have
checked that those splitted states due to inter-species interaction has similar
feature in terms of wave function symmetry. In addition,
when spin-up exciton condensate is dragged in momentum space from $-\boldsymbol{K}_m$ to $\boldsymbol{\Gamma}$, spin-down exciton condensate will be dragged from
$\boldsymbol{K}_m$ to $-\boldsymbol{K}_m$ accordingly, which remains dark during the whole process.
So our prediction above on optical selection rules and
path dependent light polarization remains largely intact.

In summary, we have studied the properties of moi\'e exciton condensate with exciton-exciton interaction, focusing on the
nonlinear Bloch band and Bloch waves.
The nonlinear dispersion features loop or swallowtail structures, with an emergent nonlinear
Dirac cone at $\boldsymbol{\Gamma}$ point. There is three-fold degeneracy at nonlinear
Dirac point, and each degenerate Bloch wave breaks $C3$ rotational
symmetry. The symmetry change of Bloch waves at
$\boldsymbol{\Gamma}$ point has dramatic effect on their optical selection rules, changing from circular polarization to polarization with all three components. 
Those three degenerate states can be individually reached by evolving smoothly from
 $-\boldsymbol{K}_m$ to $\boldsymbol{\Gamma}$ along three different paths. During the evolution process, the Berry curvature of lowest Bloch band is significantly reduced compared with
non-interacting case.
 These properties allow
us to propose an experimental scheme to verify our predictions. By applying an
in-plane electric field only on one layer, the electric field can drag an
exciton condensate from $-\boldsymbol{K}_m$ to $\boldsymbol{\Gamma}$ in moir\'e mini-BZ.
One will experimentally find that the photoluminescence at $\boldsymbol{\Gamma}$
has all three components of polarization, which is dependent on
the direction of applied electric field. Besides,
the signal of Hall effect induced by Berry curvature of nonlinear Bloch band is much weaker
than the single exciton case. 

\section{acknowledgments}

We are grateful to Yongping Zhang, Hongyi Yu and Wang Yao for valuable discussions. This work is
supported by the National Natural Science Foundation of China (Grant No.
12004118), the Guangdong Basic and Applied Basic Research Foundation (Grants
No. 2020A1515110228 and No. 2021A1515010212), and the Science and Technology
Program of Guangzhou (Grant No. 2019050001).

\bibliographystyle{apsrev4-2}

\end{document}